\documentclass[
11pt,showpacs,preprint,tightenlines,amsmath,amssymb,superscriptaddress]{revtex4}
\usepackage{graphicx}
\usepackage{dcolumn}
\usepackage{bm}
\usepackage{amsmath}
\usepackage{psfig}
\usepackage{epsfig}

\def\beq{\begin{equation}}
\def\eeq{\end{equation}}
\def\bea{\begin{eqnarray}}
\def\eea{\end{eqnarray}}
\def\beqa{\begin{equation}\begin{array}{l}}
\def\eeqa{\end{array}\end{equation}}
\def\eqlab#1{\label{eq:#1}}
\def\figlab#1{\label{fig:#1}}

\def\eref#1{(\ref{eq:#1})}
\def\Eqref#1{Eq.~(\ref{eq:#1})}

\def\Figref#1{Fig.~\ref{fig:#1}}

\def\half{\mbox{\small{$\frac{1}{2}$}}}

\def\barr{\left(\begin{array}{c}}
\def\earr{\end{array}\right)}
\def\bmat{\left(\begin{array}{cc}}
\def\emat{\end{array}\right)}
\def\al{\alpha}
\def\be{\beta}
\def\ga{\gamma} 
\def\de{\delta} 
\def\veps{\varepsilon}  

\def\la{\lambda}

\def\si{\sigma} 
\def\th{\theta}  \def\Th{\Theta}
\def\w{\omega}  
\def\vfi{\varphi}

\def\bq{{\bf q}} 
\def\bp{{\bf p}}

\def\mathscr{\mathcal}

\def\3d{3-D}

\def\eqlab#1{\label{eq:#1}}
\def\figlab#1{\label{fig:#1}}

\def\eref#1{(\ref{eq:#1})}
\def\Eqref#1{Eq.~(\ref{eq:#1})}

\def\Figref#1{Fig.~\ref{fig:#1}}

\begin{document}
\title{Solving Potential Scattering Equations without Partial Wave Decomposition}

\author{George Caia}
\email{caia@phy.ohiou.edu} \affiliation{ Institute of Nuclear and
Particle Physics (INPP), Department of Physics and Astronomy,
Ohio University, Athens, OH 45701}
\author{Vladimir Pascalutsa}
\email{vlad@jlab.org} \affiliation{ Institute of Nuclear and
Particle Physics  (INPP), Department of Physics and Astronomy,
Ohio University, Athens, OH 45701}
\affiliation{Department of Physics, College of William \& Mary, Williamsburg, VA 23188 \\
{\em and} Theory Group, JLab, 12000 Jefferson Ave, Newport News,
VA 23606}
\author{Louis E. Wright}
\email{wright@phy.ohiou.edu} \affiliation{ Institute of Nuclear
and Particle Physics (INPP), Department of Physics and Astronomy,
Ohio University, Athens, OH 45701}

\begin{abstract}
Considering two-body integral equations we show how they can be
dimensionally reduced by integrating exactly over the azimuthal
angle of the intermediate momentum. Numerical solution of the
resulting equation is feasible without employing a partial-wave
expansion. We illustrate this procedure for the Bethe-Salpeter
equation for pion-nucleon scattering  and  give explicit details
for the one-nucleon-exchange term in the potential.  Finally, we
show how this method can be applied to pion photoproduction from
the nucleon  with $\pi N$ rescattering being treated so as to
maintain unitarity  to first order in the electromagnetic
coupling. The procedure for removing the azimuthal angle
dependence becomes increasingly complex as the spin of the
particles involved increases.
\end{abstract}

\maketitle

\section{Introduction}

In cases when solving the Lippmann-Schwinger or Bethe-Salpeter
type of equation is numerically involved, one often resorts to a
partial-wave decomposition (PWD) in the center-of-mass (CM)
frame. In doing so one can exploit the spherical symmetry of the
interaction and perform the integration over the two-dimensional
solid angle of the intermediate momentum analytically. While this
reduces the equation's dimension by two, one has to deal with
summing the partial-wave series, and hence this procedure is
beneficial when only a few partial waves dominate. In the case
when many partial waves must be taken into account, when
restriction to the CM frame is not desirable, or when the
potential is not spherically-symmetric, the partial-wave
expansion is not helpful and one has to face the complexity of
three- or four-dimensional integral equations.

Fortunately, as had been noted by Gl\"ockle and
collaborators~\cite{Glo88,Elster} in the context of the
nucleon-nucleon ($NN$) interaction, the dependence on the
intermediate momentum azimuthal angle factorizes and can still be
performed analytically without employing any kind of expansion or
truncation. While this procedure has  been successfully applied a
number of times to the $NN$ situation~\cite{Elster,DeW92,RAP00},
here we would like to examine general conditions which potentials
must satisfy to factorize the azimuthal integration. We then
apply it to solve a specific example of relativistic potential
scattering in the pion-nucleon ($\pi N$) system and compare with
the usual method of using the partial-wave expansion.

In Section II we give the general requirements on the potential
that allow one to remove the azimuthal angle dependence in the
integral equation.  In Section III we focus on the Bethe-Salpeter
equation for $\pi$N scattering with one-nucleon-exchange
potential and show in detail how the azimuthal-angle dependence
can be integrated out in this case. Furthermore, in Section IV,
we solve the resulting equation using a quasipotential
approximation and compare the solution to the one obtained using
the partial-wave expansion. In Section V we examine an extension
of this approach to the calculation of  pion electroproduction
from the nucleon including the $\pi N$ final state interaction.
Our conclusions are summarized in Section VI.

\section{Conditions for exact integration over the azimuthal angle}
The starting point in calculating observables of a two-body
scattering process is an equation for the scattering amplitude.
We shall assume relativistic scattering, in which case the
equation is a 4-dimensional integral equation of the
Bethe-Salpeter type:
\begin{eqnarray}
\eqlab{BS4-D}
 T(q',q;P) =  V(q',q;P)  +i\int \frac{d^4 q''}{(2\pi)^4}\, V(q',q'';P)\,
G(q'';P)\,T(q'',q;P)
\end{eqnarray}
where $T$ is the sought T-matrix, $G$ is the two-particle
propagator, and $V$ is the two-particle-irreducible potential.
Moreover, throughout the paper, ${q}$, $q''$, $q'$ stand for the
relative 4-momenta of the incoming/intemediate/outgoing channel
while $P=p+k=p'+k'=p''+k''$ is the total 4-momentum with $k$,
$k''$, $k'$ and $p$, $p''$,  $p'$ the
incoming/intermediate/outgoing momenta of particle one and
particle two, respectively.

In order to investigate the conditions under which the above
equation can be integrated over the intermediate azimuthal angle
we work in the  helicity basis and only display the dependence on
the azimuthal angle and helicity:
\begin{eqnarray}
 T_{\la'\la}(\vfi',\vfi) = V_{\la'\la}(\vfi',\vfi)   +\sum_{\la''}\int\limits_0^{2\pi} \!\frac{d\vfi''}{2\pi}
 V_{\la'\la''}(\vfi',\vfi'')\, G(\vfi'')\, T_{\la''\la}(\vfi'',\vfi)\, .
\eqlab{eqn} \end{eqnarray}

An important point here is that the two-particle propagator $G$
can always be made independent of the intermediate angle $\vfi''$
by choosing the total three-momentum along the $z$-axis, i.e.,
choosing the {\em co-linear frame}: $P=(P_0,0,0,P_3)$.
Furthermore, we shall observe that in the case when only spin-0
and spin-1/2 particles are involved, the azimuthal-angle
dependence of the fully off-shell potential \footnote { In
general, we deal with the fully off-shell situation, that is when
both initial and final states are off the mass (or energy, in the
non-relativistic case) shell.   The situation when either the
initial or the final state is on-shell is referred to as the
half-off-shell case, and it is well known that one only needs the
half-off-shell result to solve the integral equation.} in the
co-linear frame is given as follows
\begin{eqnarray}
V_{\la'\la}(\varphi',\varphi) = e^{-i\la'\vfi'}\,
v_{\la'\la}(\varphi'-\varphi)\, e^{i\la\vfi} \eqlab{2a}
\end{eqnarray}
where $\la$ and $\la'$ stand for the combined helicities of the
initial and final state, respectively. The half-off-shell
potential then takes a very simple form: \beq \left.
V_{\la'\la}(\varphi',\varphi) \right|_{{\rm half-off-shell}}
=\left. \,e^{-i(\la'-\la)\vfi'} v_{\la'\la}(0)\right|_{{\rm
half-off-shell}}\eqlab{2b} \eeq where $\la$ is the helicity of
the on-shell state.

It is in this case, when conditions \eref{2a} and \eref{2b} are
met, the exact integration over the azimuthal-angle can readily
be done. First, by using \eref{2a} in \Eqref{eqn}, we see that
the azimuthal dependence of the t-matrix is given by: \beq
T_{\la'\la}(\varphi',\varphi) =
e^{-i\la'\vfi'}\,t_{\la'\la}(\varphi'-\varphi)\, e^{i\la\vfi}.
\eqlab{eq3} \eeq Since $v$ and $t$ only depend on difference
$\vfi'-\vfi$, we expand them in a simple Fourier series: \beq
v_{\la'\la}(\phi) = \sum_m v^{(m)}_{\la'\la} \, e^{im\phi},\,\,\,
\hspace{0.1cm} t_{\la'\la}(\phi) = \sum_m t^{(m)}_{\la'\la} \,
e^{im\phi}. \eeq It is straightforward to show that their Fourier
transforms, \beq v^{(m)}_{\la'\la} = \int\limits_0^{2\pi}
\!\frac{d\phi}{2\pi} \,v_{\la'\la}(\phi)\, e^{-i m
\phi},\,\,\,\hspace{0.05cm} t^{(m)}_{\la'\la} =
\int\limits_0^{2\pi} \!\frac{d\phi}{2\pi} \,t_{\la'\la}(\phi)\,
e^{-i m \phi}\,, \eeq satisfy the following equation which  does
not involve the $\vfi$-integration: \beq t^{(m)}_{\la'\la} =
v^{(m)}_{\la'\la} + \sum_{\la''} v^{(m)}_{\la'\la''}\,G\,
t^{(m)}_{\la''\la}\,. \eqlab{eqn2} \eeq

In principle, $m$ runs to infinity and so we have an infinite
number of equations to solve even though they are not coupled.
Fortunately, since only the half off-shell potential is needed to
solve the equations and it obeys condition \eref{2b}, the
corresponding Fourier transform is non-vanishing only for
$m=-\la$: \beq \left.v^{(m)}_{\la'\la}\right|_{{\rm
half-off-shell}} =  \left.\delta_{-\la m}
\,v_{\la'\la}(0)\right|_{{\rm half-off-shell}}. \eeq

The scalar system is the simplest one where this procedure can be
demonstrated. In that case the potential is a scalar function of
scalar products of relevant 4-momenta: \beq \eqlab{scalar}
V(q',q; P)=V(q\cdot q', P\cdot q, P\cdot q', q^2, {q'}^2, P^2)
\eeq Given  $q=(q_0,\, |\bq|\, \sin\th\, \cos\vfi,\, |\bq|
\sin\th\,\sin\vfi,\, |\bq|\, \cos\th)$ and similarly for $q'$, we
easily convince ourselves that, in the co-linear frame, the
azimuthal dependence enters only through the product: \beq q\cdot
q'  = q_0 q_0' - |\bq|\,|\bq'|\,\left[\cos\th \cos\th' + \sin\th
\sin\th'\, \cos(\vfi'-\vfi)\right] \eeq and hence it is of the
necessary  form given in \Eqref{2a}. Furthermore, in the
half-off-shell case the momentum of the on-shell state, say $q$,
can always be chosen along the $z$-axis, i.e., such that $\th=0$.
Hence the half-off-shell potential is independent of azimuthal
angles which fulfills condition \eref{2b} for the spinless case.
The two-particle propagator $G(q;P)=G(P\cdot q, q^2, P^2) $ is of
course independent of $\vfi$ in the co-linear frame.

Once we have found that conditions \eref{2a} and \eref{2b} are
satisfied, while $G$ is independent of $\vfi$, the integration
over $\vfi$ can be done immediately. We will now show this more
explicitly for the more complicated case of a scalar-spinor
system.

\section{Spin complications: the $\pi N$ system}
Consider the Bethe-Salpeter equation for the case of elastic
scattering of a scalar with mass $m_\pi$ --- the ``pion'' --- on
a spinor with mass $m_N$ --- the ``nucleon''. We attribute the
momenta $p$, $p'$ to the nucleon and $k$, $k'$ to the pion. The
relative 4-momentum of the incoming channel is conveniently
defined by $q=\beta p-\alpha k$,  where Lorentz scalars $\al$ and
$\be$ are given by 
\bea
\eqlab{albe}
\alpha &=& p\cdot P/s= (s+m_N^2-m_\pi)/2s\,,\nonumber\\
\beta &=&k\cdot P/s=(s-m_N^2+m_\pi)/2s\,, \eea with $s=P^2$.
Similarly one defines $q'=\beta p'-\alpha k'$ and
  $q''=\beta p''-\alpha k''$ as the relative
 4-momenta of the outgoing and intermediate state, respectively.
In terms of these variables,
  the two-body $\pi$N Green's function of \Eqref{BS4-D} is:
\begin{eqnarray}
\eqlab{propag} G(q;P)=\frac{1}{(\beta P-q)^{2}-m_{\pi
}^{2}+i\epsilon} \frac{(\alpha P+q)\cdot
 \gamma +m_{N}}{(\alpha P+q)^{2}-m^{2}_{N}+i\epsilon }.
\end{eqnarray}
\begin{figure}[t]
\epsffile{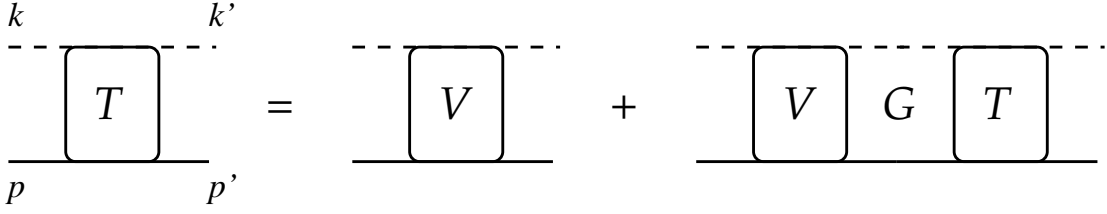} \caption{Diagrammatic form of a relativistic
two-body scattering equation.} \figlab{bsef1}
\end{figure}
Projecting the equation onto the basis of the nucleon helicity
spinors (defined in Appendix A), we obtain
\begin{eqnarray}
\eqlab{BetSalproj} T_{\lambda'\lambda}^{\rho'\rho}(q',q;P)  =
V^{\rho'\rho}_{\lambda'\lambda}(q',q;P)
 + i\sum_{\lambda '' \rho ''}\int \!\frac{d^4 q''}{4\pi^{3} }V^{\rho'\rho''}_{\lambda'\lambda''}
 (q',q'';P)\, G^{(\rho'')}(q'';P)\,T_{\lambda''\lambda}^{\rho''\rho}(q'',q;P),
\end{eqnarray}
where the helicity amplitudes are defined as \beq T^{\rho'
\rho}_{\lambda' \lambda}(q',q,P)=(1/4\pi)\, \bar
u^{(\rho')}_{\lambda'}(\al {\bf P}+{\bf q'}) \,\,T(q',q,P)\,\,
u^{(\rho)}_{\lambda}(\al {\bf P}+{\bf q}), \eeq
 and analogously for $V$,
while the defining equation for $G^{(\rho)}$ is \beq
\bar{u}^{(\rho')}_{\lambda'}(\al {\bf P}+{\bf
q})\,\gamma^{0}\,G(q;P)\,\gamma^{0}\,u^{(\rho )}_{\lambda }(\al
{\bf P}+{\bf q})=\delta _{\lambda'\lambda}\,\delta _{\rho '\rho
}\,G^{(\rho)}(q;P), \eeq and hence \beq \eqlab{gpm}
G^{(\pm)}(q;P)=\frac{1}{q_0+\alpha \sqrt{s}\pm (E_{\al P
+q}-i\epsilon)} \frac{1}{(\beta \sqrt{s}-q_0 )^{2}-\w_{\be P
-q}^{2}+i\epsilon}, \eeq with $E_{q}=\sqrt{{\bf q}^{2}+m_N^{2}}$
and $\w_{q}=\sqrt{{\bf q}^{2}+m_\pi^{2}}$.

The most general Lorentz structure of the fully off-shell
potential in the helicity basis can be written in the form
\footnote{To bring a general expression to this form we use
properties of the Dirac spinors, such as:
$$
(\gamma \cdot q - m_N) \,u_{\lambda}^{\rho}({\bf q})= (q_{0}-\rho
E_{q})\, \gamma^{0}\,u_{\lambda}^{\rho}({\bf q}).
$$}
\beq \eqlab{potproj} V^{\rho'\rho}_{\lambda' \lambda ''}(q',q;P) =
\bar{u}^{\rho'}_{\lambda'}(\al {\bf P}+{\bf q}')\, \left[
A_{1}^{\rho'\rho} +A^{\rho'\rho}_{2}\,\gamma^{0} +
(A^{\rho'\rho}_{3} + A^{\rho'\rho}_{4}\,\gamma^{0}) \,{\bf
\gamma} \cdot {\bf P} \right] \,u^{\rho}_{\lambda }(\al {\bf
P}+{\bf q}), \eeq where $A_i$ are scalar functions of the
dot-products of the relevant momenta, i.e., \beq
 A_i = A_i(q\cdot q', P\cdot q, P\cdot q', q^2, {q'}^2, P^2).
\eeq Considering the dependence of these functions on the
azimuthal angles of $q$ and $q'$, we see  that --- in the {\em
co-linear frame} --- it is given by the difference $\vfi'-\vfi$,
for the reason described below \Eqref{scalar}.

The rest of the $\vfi$-dependence resides in the nucleon spinors.
According to \Eqref{potproj}, in the co-linear frame we need to
consider only $  \chi^\dagger_{\la'} (\Theta',\vfi') \,
\chi_\la(\Theta,\vfi) \,\,\,\, \mbox{and}\,\,\,\,
 \chi^\dagger_{\la'} (\Theta',\vfi') \,\si_3\, \chi_\la(\Theta,\vfi)
$ where $\chi$'s are the Pauli spinors (cf.~Appendix~A),
$\Theta$, $\vfi$ and $\Theta'$, $\vfi'$ define the orientation of
$\al{\bf P}+\bq$ and  $\al{\bf P}+\bq'$, respectively. Since, \bea
\chi^\dagger_{\la'} (\Theta',\vfi') \, \chi_\la(\Theta,\vfi) &=&
e^{-i\la'\vfi'}\left[\sum\nolimits_{\la''}
d^{1/2}_{\la'\la''}(\Theta')\,
d^{1/2}_{\la\la''}(\Theta) e^{i\la''(\vfi'-\vfi)}\right] e^{i\la\vfi}, \\
\chi^\dagger_{\la'} (\Theta',\vfi') \,\si_3\,
\chi_\la(\Theta,\vfi) &=&
e^{-i\la'\vfi'}\left[\sum\nolimits_{\la''}(-1)^{1/2-\la''}\,
d^{1/2}_{\la'\la''}(\Theta')\, d^{1/2}_{\la\la''}(\Theta)
e^{i\la''(\vfi'-\vfi)}\right] e^{i\la\vfi}, \eea we observe that
the $\vfi$-dependence of these elements is of the desired
form~\Eqref{2a}. And for the half-off-shell situation, where we
can choose $\th=0$ (hence $\Th=0$, in the co-linear frame) and use
$d^{1/2}_{\la\la''}(0)=\de_{\la\la''}$, we find the form, \bea
\chi^\dagger_{\la'} (\Theta',\vfi') \, \chi_\la(0,\vfi) &=&
e^{-i(\la'-\la)\vfi'} d^{1/2}_{\la'\la}(\Theta'), \\
\chi^\dagger_{\la'} (\Theta',\vfi') \,\si_3\, \chi_\la(0,\vfi) &=&
e^{-i(\la'-\la)\vfi'}\,(-1)^{1/2-\la}\,
d^{1/2}_{\la'\la}(\Theta'), \eea which obeys the necessary
half-shell condition \Eqref{2b}.

Therefore, we have demonstrated that the azimuthal-angle
dependence of a pion-nucleon potential {in the co-linear frame}
always satisfies conditions  \eref{2a} and \eref{2b}. It is also
apparent from \Eqref{gpm} that the two-particle Green's function
does not have any azimuthal dependence in that frame. Thus the
integration over $\vfi$ can exactly be done in the Bethe-Salpeter
equation for $\pi N$ system by means of the procedure of Sec.~II.

Similar arguments apply in the case when both particles have spin
1/2, {\em e.g.}, the nucleon-nucleon (NN) scattering. It should
only be noted that in this case the potential  satisfies
conditions  \eref{2a} and \eref{2b} with $\la=\la_1-\la_2$,
$\la'=\la_1'-\la_2'$. In other words, helicities of the two
particles must be combined.


\section{Numerical results}

The standard route to solution of a potential scattering equation
such as \Eqref{BetSalproj} is to decompose it into an infinite
set of equations for partial-wave  amplitudes, see
e.g.~\cite{JaW59,Kub72}. The advantage of doing a partial wave
decomposition  is that the equation for each partial wave is of 2
lesser dimensions than the original equation, while the
partial-wave series is usually rapidly converging, hence only the
first few partial-wave amplitudes need to be solved for.

On the other hand, solving for the full amplitude directly has its
own important benefits. And if the exact azimuthal-angle
integration can  be done {\it a priori}, the numerical
feasibility of this approach becomes comparable to the PWD method.
\begin{figure}[t]
\epsffile{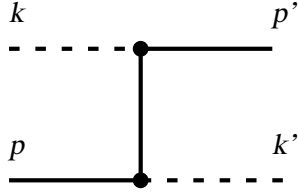} \caption{One-nucleon-exchange $\pi N$
potential.} \figlab{uchan2}
\end{figure}

In this section we would like to compare the two methods for the
example of solving a relativistic equation for the $\pi N$
system. For our toy-calculation potential we take the one-nucleon
exchange, \Figref{uchan2}, and use the {\it instantaneous}
approximation, thus neglecting retardation effects in the
potential. The latter approximation allows us to perform the
relative-energy ($q_0$) integration such that we are left with a
relativistic 3-dimensional Salpeter equation: \beq
T_{\lambda'\lambda}^{\rho'\rho}(\bq',\bq;P)  =
V^{\rho'\rho}_{\lambda'\lambda}(\bq',\bq;P) + \sum_{\lambda ''
\rho ''}\int \!\frac{d^3 q''}{4\pi^{2}
}V^{\rho'\rho''}_{\lambda'\lambda''}
 (\bq',\bq'';P)\, G_{ET}^{(\rho'')}(\bq'';P)\,T_{\lambda''\lambda}^{\rho''\rho}(\bq'',\bq; P),
\eeq where the equal-time two-particle propagator in the CM
system is given by
\begin{eqnarray}
G^{(\rho) }_{ET}(|{\bf q}|;\sqrt{s})
 = 2i\int _{-\infty }^{\infty }\frac{dq_{0}}{2\pi }\,G^{(\rho) }(q;P)
=\frac{-\rho}{\omega _{q}(-\rho \sqrt{s}+E_{q}+\omega
_{q}-i\epsilon ) }
\end{eqnarray}
This 3-dimensional equation for $\pi N$ has been described in
detail and solved using a PWD in the CM system by Pascalutsa and
Tjon~\cite{PaT98, Pas98,PaT00}. We, on the other hand, solve this
equation by using the framework of the two previous sections to
reduce the $\vfi$-integration analytically and solve numerically
the resulting 2-dimensional integral equation for the $m$-th
Fourier component of the full amplitude:
\begin{eqnarray}
\eqlab{BetSal2D} && t_{\lambda'\lambda}^{(m)\rho'\rho}(|{\bf
q'}|,\th',|{\bf q}|,\theta)  =
{v}_{\lambda'\lambda,}^{(m)\rho'\rho}(|{\bf q'}|,\th', |{\bf q}|,\theta)\nonumber \\
 && +\sum_{\lambda'' \rho''}\int \nolimits_{0}^{\infty} \frac{d|\bq''|}{2\pi}\,
|\bq''|^{2}\int\nolimits _{0}^{\pi } d\theta''
v_{\lambda''\lambda'}^{(m) \rho''\rho'}(|{\bf q'}|,\th',|{\bf
q''}|,\theta'')\, G_{ET}^{(\rho'')}(|{\bf q'' }|) \,
t_{\lambda''\lambda}^{(m) \rho''\rho}(|{\bf q''}|,\th'',|{\bf
q}|,\theta),
\end{eqnarray}
where, without loss of generality, we have also assumed the CM
frame. The explicit form of the Fourier transform of the
one-nucleon-exchange potential is worked out  in Appendix B.

Let us emphasize that it is necessary to solve for only one of
the Fourier components (either $m=-1/2$ or $m=1/2$), the other
ones either vanish or can be obtained by relations due to the
parity and time-reversal invariance.

We solve \Eqref{BetSal2D} by the Pade approximants as in
Refs.~\cite{Pas98,PaT00} thus maintaining exact elastic
unitarity. The numerical integrations are performed by the
Gauss-Legendre method. The integral over $|{\bf q''}|$ in
\Eqref{BetSal2D} contains the cut singularity at $|{\bf q''}|
=\sqrt{[s-(m_{N}-m_{\pi})^{2}][s-(m_{N}+m_{\pi})^{2}]/4s}\equiv
\hat q$, which is handled by the well-known identity:
\begin{eqnarray}
\eqlab{prin} \int\nolimits_{0}^{\infty}d|\bq| \frac{f(|\bq|)}
{|\bq|-\hat q +i\epsilon}={\cal P}\int\nolimits_{0}^{\infty}d|\bq|
\frac{f(|\bq|)}{|\bq|-\hat q }-i\pi f(\hat q),
\end{eqnarray}
where $\cal P$ denoted the principal-value integral. When
computing the latter the integration region is divided into two
intervals: $|{\bf q}| \in [0,2\hat q]$, and  $|{\bf q}| \in
(2\hat q,\infty )$. The Gaussian  points are then distributed
separately for each interval to  make use of the property that an
even number of Gaussian points falls  symmetrically with respect
to the middle of the interval hence the singularity in the middle
of the first interval is avoided. The polar angle integration is
straightforward for both the principal value term and the
imaginary contribution. We find it sufficient to use 16 Gauss
points for the momentum integration and 8 points for the
polar-angle integration.  Upon increasing the number of points to
32 and 16 respectively, the results change by less than $0.5\%$
in the considered energy range. In all cases we found that 6
iterations combined with the use of Pad\`{e} approximants works
extremely well.

\begin{figure}[th]
\begin{center}
\includegraphics[totalheight=0.32\textheight]{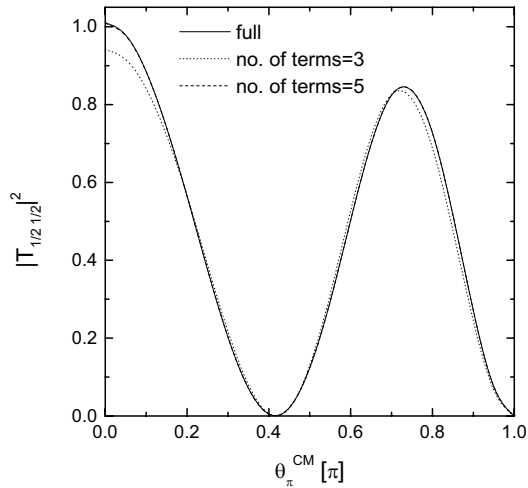}
\caption{\label{angle300f3} Angular dependence for
$|T_{\frac{1}{2}\frac{1}{2}}^{+ +}|^{2}$ at $E_{\pi }^{LAB}=300$
MeV. Solid line is the full calculation, dashed and dotted are
the resumming of partial terms.}
\end{center}
\end{figure}

\begin{figure}[ht]
\begin{center}
\includegraphics[totalheight=0.32\textheight]{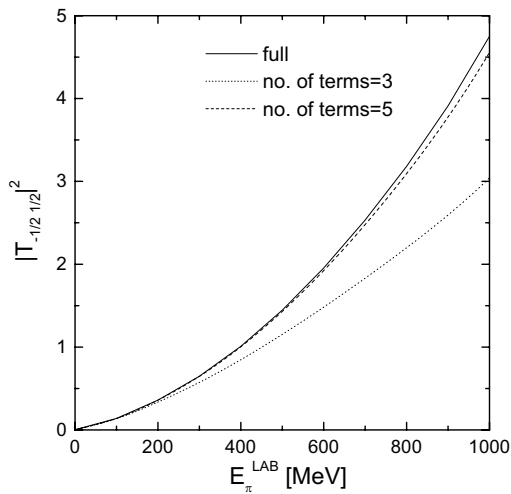}
\caption{\label{energy1PIf4} Energy dependence for
$|T_{-\frac{1}{2}\frac{1}{2}}^{+ +}|^{2}$ at $\theta_{\pi
}^{CM}=\pi $. The lines are defined the same as in
Fig.\ref{angle300f3}}
\end{center}
\end{figure}

After we solve \Eqref{BetSal2D} to find the full $\pi N$
$T$-matrix,  we  can of course also find the partial wave
amplitudes:
\begin{eqnarray}
\eqlab{pwT} T^{J\rho '\rho }_{\lambda '\lambda
}(|\bq'|,|\bq|)=\int_{0}^{\pi}\!d\theta \,T_{\lambda '\lambda
}^{\rho '\rho}(|\bq'|,|\bq|,\th)\, d_{\lambda '\lambda
}^{J}(\theta ),
\end{eqnarray}
where $\theta$ is the  angle between ${\bf q}$ and ${\bf q'}$. We
then investigate the convergence of the partial wave series:
\begin{eqnarray}
\eqlab{pwd} T_{\lambda '\lambda }^{\rho '\rho}(|\bq'|,|\bq|,\th)=
\sum_{J}\left(J+\half\right)\, T^{J\rho '\rho }_{\lambda '\lambda
}(|\bq'|,|\bq|)\, d_{\lambda '\lambda }^{J}(\theta )\,.
\end{eqnarray}
In particular,  in Fig.\ref{angle300f3} and Fig.\ref{energy1PIf4}
we plot the on-shell values of  $|T_{\lambda '\lambda }^{\rho
'\rho }|^{2}$ compared with the truncation of the partial-wave
series for 3 terms and 5  terms (i.e.,
$J=\frac{1}{2},\ldots,\frac{5}{2}$ and
$J=\frac{1}{2},\ldots,\frac{9}{2}$ respectively).

In order to compare the computational efficiency of the two
methods, we compare the number of partial waves needed to achieve
convergence in the PWD method with the number of Gauss points for
the polar-angle integration which appear in the ``w/o PWD''
method.

The figures show that the effect of  truncations of the
partial-wave series increases with the angle
(Fig.~\ref{angle300f3}) the energy of the incoming $\pi$
(Fig.~\ref{energy1PIf4}). In our particular case of one-nucleon
exchange computing $5$ or more partial wave amplitudes is
sufficient to reproduce the full result to a 1 per cent accuracy
in a broad energy domain. Thus, in this case, the efficiency of
the two methods is comparable since we need 5 multipoles versus 8
Gauss points of the polar-angle integration.

It is important to emphasize that the ability to do the
azimuthal-angle integration analytically is necessary to achieve
comparable efficiency. We have checked that it usually takes at
least 16 Gaussian points for the azimuthal integration which
slows down the calculation by more than an order of magnitude.

\section{Extension to  pion photoproduction}
Our procedure for performing the analytic $\vfi$-integration is
applicable in the photo- or electro-meson production to first
order in the electromagnetic coupling.
 Here we describe the extension to the case of $\pi$ photoproduction
within a simple final-state-interaction model~\cite{Pas98,PaT03}.
 The model begins with the following coupled channel
 equation:
\beq \left( \begin{array}{cc}
T_{\pi \pi } & T_{\pi \gamma }  \\
T_{\gamma  \pi } & T_{\gamma \gamma } \end{array} \right)=\left(
\begin{array}{cc}
V_{\pi \pi } & V_{\pi \gamma }  \\
V_{\gamma  \pi } & V_{\gamma \gamma } \end{array} \right)+\left(
\begin{array}{cc}
V_{\pi \pi } & V_{\pi \gamma }  \\
V_{\gamma  \pi } & V_{\gamma \gamma } \end{array} \right)\left(
\begin{array}{cc}
G_{\pi} & 0 \\
0 & G_{\gamma } \end{array} \right)\left( \begin{array}{cc}
T_{\pi \pi } & T_{\pi \gamma }  \\
T_{\gamma  \pi } & T_{\gamma \gamma } \end{array} \right) \eeq
where $T$ and $V$ are the amplitudes and driving potentials of
the $\pi$N scattering $(\pi \pi)$, pion photo-production $(\gamma
\pi)$, absorption $(\pi \gamma)$, and the nucleon Compton effect
$(\gamma \gamma)$, respectively. The above equations are solved
up to first order in the electromagnetic coupling $e$, hence
preserving two body unitarity to this order only.

In solving the photoproduction scattering equation we calculate
first $V_{\pi \pi }$ as described for $\pi$N scattering and we
then  iterate in the following manner:
\begin{eqnarray}
\eqlab{pigaiter} T_{\pi \gamma }=V_{\pi \gamma }+V_{\pi \pi
}G_{\pi }V_{\pi \gamma }+ V_{\pi \pi } G_{\pi }V_{\pi \pi }G_{\pi
}V_{\pi \gamma }+\ldots,
\end{eqnarray}
where we used $T_{\pi \gamma}=T_{\gamma \pi }$ from time-reversal
invariance.

This solution procedure is obviously suitable for our case since
the half-shell $V_{\pi \gamma }$  has a simple azimuthal angle
dependence similar to the case of $V_{\pi \pi }$ (see
\Eqref{halfpot}). The reduced kernel (see \Eqref{reducedv}) has
two terms rather than the one term in the $\pi$N case due to the
''complication" of having to couple a spin 1 photon to spin 1/2
as opposed to coupling a spin 0 meson to spin 1/2.  For example,
if one considers the nucleon u-channel exchange (compare to the
$\pi$N case in \Eqref{potprojCM}) the half-shell photoproduction
potential can be written as: \beq \eqlab{potpiga} V_{\lambda
'\lambda \sigma }^{\rho '\rho }(q',q)= {}_1V_{\lambda '\lambda
\sigma }^{\rho '\rho }(q'_{0},|{\bf q'},q_{0},|{\bf q}|,\theta
')e^{-i(\lambda '-\lambda -\sigma )\phi '} + {}_2V_{\lambda
'\lambda \sigma}^{\rho '\rho }(q'_{0},|{\bf q'},q_{0},|{\bf
q}|,\theta ')\, e^{-i(\lambda '+\lambda)\phi '} \eeq where
$\sigma =\pm 1$ represents the helicity of the incoming photon.

One sees that when \Eqref{potpiga} is iterated in
\Eqref{pigaiter} two de-coupled scattering equations are obtained
(each corresponding to ${}_1V$ or ${}_2V$).  For each of these
equations, one can show that the corresponding $\vfi'$ dependence
re-appears after doing the $\vfi''$ integration and therefore
once again we can perform the azimuthal-angle integration
analytically. As in the $\pi$N case, the resulting ''reduced''
kernels obey  2-D integral equations.

As a check of our procedures we calculated the $u$-channel
contribution to pion photoproduction using the analytic
azimuthal-angle
 integration along with 2-D numerical integration and
compared to the results of Refs.~\cite{Pas98,PaT03} obtained
using the multipole expansion. At $E_{\gamma}$ of $300$ MeV with
five multipoles we found agreement to better than 1\% over a wide
angular range.

\section{Conclusion}

In recent years Gl\"ockle and collaborators~\cite{Glo88,Elster}  introduced a
method which greatly simplifies the numerical integration of 
 two-body scattering equations without performing the partial-wave expansion. 
The method exploits a certain azimuthal symmetry of the potential thus allowing
exact integration of the azimuthal dependence.  In this paper
we have established general form of the azimuthal-dependence of the
kernel  which allows for this procedure to go through.
We have argued that these conditions is in general applicable to any
system of spin- 0 and/or spin- $\frac{1}{2}$ particles

We have applied this
method to the case of $\pi$N system and . With some extra effort it can be
applied to higher spin systems, however the procedure becomed increasingly complex
with the increase of the spin of the involved particles. We have
successfully applied the method to pion
photo- and electro-production from the nucleon, however only to the
leading order in electromagnetic coupling.

Even though we have used the Salpeter equation for or numerical
exercises, the method can of course be applied to the full 4-D
Bethe-Salpeter equation, which for the $\pi N$ system has so far
been solved in partial waves only~\cite{NiT68, Lah99}.
Performing the azimuthal-angle integration analytically
greatly facilitates  finding the full solution and makes the numerical
feasibility of this approach comparable to finging the solution
using the  partial-wave expansion.

\appendix
\section{Helicity spinors}
We define the four-component nucleon helicity spinors as follows:
\beq
u_{\la}(E_p,{\bf p})= \frac{1}{\sqrt{2E_p}}\left( \begin{array}{c} \sqrt{E_p+m_N} \\
                         2\la \sqrt{E_p-m_N}\end{array} \right)
              \otimes  \chi_{\la} (\theta,\vfi)
\eeq where $\la=\pm 1/2$ is the helicity, $E_p=\sqrt{|{\bf
p}|^2+m_N^2}$ is the energy, $\theta$ and $\vfi$ are the
spherical angles of the 3-momentum ${\bf p}$, and $\chi_{\la}$ is
the two-component Pauli spinor. The positive- and negative-energy
nucleon spinors in the convention of Kubis~\cite{Kub72} are
defined as follows: \beq \eqlab{hel}
  u_{\la}^{(\pm)}({\bf p})  = u_{\la}(\pm E_p,{\bf p}) ,
\eeq They satisfy the following orthogonality and completeness
conditions: \bea && u_\la^{\dagger(\rho)} (\bp)\, u_{\la
'}^{(\rho')} (\bp)
= \delta_{\rho \rho '} \delta_{\la \la '}, \\
&&\sum_{\rho,\,\la} u_\la^{(\rho)} ( \bp)\,
u_{\la}^{\dagger(\rho)} (\bp) = 1. \eea

The Pauli spinors along the $z$-axis are given by
$$
  \chi_{1/2}(0)= \left( \begin{array}{c}
1 \\
0 \end{array} \right) , \,\,\,
  \chi_{-1/2}(0)= \left( \begin{array}{c}
0 \\
1 \end{array} \right),
$$
while along an arbitrary direction $\theta,\,\vfi$ they can be
obtained using the Wigner rotation functions:
$$
\chi_{\la}(\theta,\vfi) = \sum\nolimits_{\la'} d^{1/2}_{\la\la'}
(\theta)\,e^{i(\la-\la')\vfi}\, \chi_{\la'}(0),
$$
or, explicitly,
$$
\eqlab{chi}
 \chi_{1/2}(\theta,\vfi) = \left( \begin{array}{c}
\cos (\theta /2) \\
     e^{i\vfi}\sin (\theta /2) \end{array} \right) , \,\,\,
 \chi_{-1/2}(\theta,\vfi)  =  \left( \begin{array}{c}
                                -e^{-i\vfi}\sin (\theta /2)  \\
                                 \cos (\theta /2) \end{array} \right).
$$

\section{Azimuthal dependence of one nucleon exchange}

As an example, we consider the $u$-channel nucleon exchange
potential given by the graph in \Figref{uchan2} and the following expression,
\begin{eqnarray}
\eqlab{pot}  V(q',q;P)  =  \frac{g_{\pi NN}^2}{4m_N^2}
\gamma \cdot (\beta P-q')\gamma^{5}  
 \frac{(\al-\be)\ga\cdot P +\ga\cdot (q+q') +m_N}{[(\al-\be) P+q+q']^2-m_N^2+i\veps}
\gamma^{5} \gamma \cdot (\beta P-q)], 
\end{eqnarray}
where $\al$ and $\be$ are defined in \Eqref{albe}.
For simplicity we choose the CM frame,
where the potential in the helicity basis takes the form:
\begin{eqnarray}
\eqlab{potprojCM}
&& V^{\rho'\rho''}_{\lambda ' \lambda ''}(q',q'') =  \frac{g_{\pi NN}^2}{16\pi m_N^2}
\frac{1}{u-m_{N}^{2}}\times \nonumber \\
&&\times \bar{u}_{\rho'}^{\lambda'}({\bf
q''})[M^{\rho'\rho''}_{1}{\bf
1}+M^{\rho'\rho''}_{2}\gamma^{0}]u_{\rho''}^{\lambda''}({\bf q''})
\end{eqnarray}
with
\begin{eqnarray}
\eqlab{M1}
&& M_{1}^{\rho ' \rho ''}(p',p'')=m_{N}[2\sqrt{s}(p'_{0}+p''_{0})-s-2p'\cdotp p'' \nonumber\\
&&+p'^{2}+p''^{2}+m_{N}^{2}+(p'_{0}-\rho 'E_{p'})(p''_{0}-\rho '' E_{p''}) \nonumber\\
&&-\sqrt{s}(p'_{0}-\rho 'E_{p'}+p''_{0}-\rho '' E_{p''})]
\end{eqnarray}
\begin{eqnarray}
\eqlab{M2}
&&M_{2}^{\rho ' \rho ''}(p',p'')=  \sqrt{s}[2\sqrt{s}(p'_{0}+p''_{0})-s-2p'\cdotp p'' \nonumber\\
&&-p'^{2}-p''^{2}-3m_{N}^{2}-(p'_{0}-\rho 'E_{p'})(p''_{0}-\rho '' E_{p''})] \nonumber\\
&&+(p'^{2}+m_{N}^{2})(p''_{0}-\rho '' E_{p''}) \nonumber \\
&&+(p''^{2}+m_{N}^{2})(p'_{0}-\rho 'E_{p'}).
\end{eqnarray}

The azimuthal dependence $\phi ''$ arises from Dirac spinors and
from various scalar products involving the the four vector $q''$.
Choosing the vector part of the total momentum $P$ to be along
the $z$-axis (or to be zero in the CM frame) allows the $\phi''$
dependence, for the fully {\it off-shell} potential, to be
displayed in the form:
\begin{eqnarray}
\eqlab{potphi}
&& V^{\rho'\rho''}_{\lambda'\lambda''}(q',q'')  =   \frac{g_{\pi NN}^2}{16\pi m_N^2}N_{q'q''}{ \Omega}_{\lambda'\lambda''}(\theta',\theta'',\phi',\phi'') \nonumber \\
&& \sum_{n=0}^{N} \frac{{\cal
R}^{\rho'\rho''}_{\lambda'\lambda'',n}(q'_{0},|{\bf
q'}|,q''_{0},|{\bf q''}|,\theta',\theta'') cos^{n}(\phi''-\phi')
}{d_{1}(q'_{0},|{\bf q'}|,q''_{0},|{\bf q''}|,\theta',\theta'')+
d_{2}(q'_{0},|{\bf q'}|,q''_{0},|{\bf
q''}|,\theta',\theta'')\cos(\phi''-\phi')},
\end{eqnarray}
where ${\cal R}^{\rho'\rho''}_{\lambda'\lambda'',n}$,
$N_{p'p''}$, and $d_{i}$ are factors which depend on the type of
the diagram and of the exchanged particle, but are independent of
the azimuthal angle. The quantities,
\begin{eqnarray}
\Omega_{\lambda'
\lambda''}=e^{-i\lambda'\phi'}\sum_{m=-1/2}^{1/2}d_{\lambda'
m}^{1/2}(\theta')d_{\lambda''
m}^{1/2}(\theta'')e^{im(\phi'-\phi'')}e^{i\lambda''\phi''},
\nonumber
\end{eqnarray}
\begin{eqnarray}
N_{q'q''}=\sqrt{(E_{q'}+m_{N})(E_{q''}+m_{N})/4E_{q'}E_{q''}}
\nonumber
\end{eqnarray}
are factors which result from the helicity spinors.
 In \Eqref{potphi} we have employed the usual trigonometric relation between two arbitrary directions
  defined by ${\bf q''}$ and ${\bf q'}$:
\begin{eqnarray}
\eqlab{trig} \cos\Theta _{{\bf q''} {\bf
q'}}=\cos\theta''\cos\theta'+\sin\theta''\sin\theta'\cos(\phi''-\phi').
\end{eqnarray}

\noindent It is easy to see that the fully off-shell potential in
\Eqref{potphi} has the azimuthal dependence of \Eqref{2a}.
Furthermore, in iterating \Eqref{BetSalproj} the quantization axis
is defined by the {\it on-shell} relative momentum ${\bf q}$ (i.e.
$\theta=0$), hence \Eqref{trig} reduces to $\cos\Theta _{{\bf
q'},{\bf q}}=\cos\theta'$, therefore the {\it half-off-shell}
potential reduces to:
\begin{eqnarray}
\eqlab{halfpot}
V_{\lambda'\lambda}^{\rho'\rho}(q',q)=v_{\lambda'\lambda}^{\rho'\rho}({q_{0}',|\bf
q'}|,q_{0},|{\bf q}|, \theta')e^{-i(\lambda'-\lambda)\phi'}
\end{eqnarray}
\noindent which if of the form of the result in \Eqref{2b}.
Therefore, the azimuthal angle dependence can be removed from the
Bethe-Salpeter equation for this case.  We achieved this result by
explicitly displaying the azimuthal angle dependence and  align
${\bf P}$ with the z-axis so that only $\gamma^3$ and $\gamma^0$
appear in $V_{\lambda'\lambda}^{\rho'\rho}$.  The presence of
$\gamma^1$ or $\gamma^2$ would introduce additional azimuthal
angle dependence in the spinor matrix elements and make the
algebra much more complicated.

For the $u$-channel nucleon exchange the coefficients
$d_{i}$ are: 
\begin{eqnarray}
\eqlab{d1}
&&d_{1}(p',p'')=p'^{2}+p''^{2}+s-2 \sqrt{s}(p'_{0}+p''_{0})+2p'_{0}p''_{0}\nonumber \\
&&-2{\bf |p'|} {\bf |p''|}\cos \theta ' \cos \theta '' -m_{N}^{2}
\end{eqnarray}
\begin{eqnarray}
\eqlab{d2}
d_{2}(p',p'')=-2{\bf |p'|} {\bf |p''|}\sin \theta' \sin \theta''
\end{eqnarray}
From these relations one can exactly identify the angular
dependence of potential given in \Eqref{pot} in the 4-product,
\begin{eqnarray}
\eqlab{ppdotppp}
&&p'\cdotp p''=p'_{0}p''_{0}-{\bf p'}\cdot {\bf p''} \nonumber \\
&&=p'_{0}p''_{0}-{\bf |p'|} {\bf |p''|}\cos \theta ' \cos \theta '' \nonumber \\
&&-{\bf |p'|} {\bf |p''|}\sin \theta' \sin \theta'' \cos (\phi
'-\phi'').
\end{eqnarray}
The relative momenta $q'$ and $q''$, defined in SECTION II, are
to be introduced in \Eqref{M1} - \Eqref{ppdotppp} by $p'=\alpha
P+q'$ and $p''=\alpha P+q''$

where
\begin{eqnarray}
\eqlab{reducedv}
 \bar{v}_{\lambda''\lambda',\lambda}^{\rho''\rho'}(|{\bf q'}|,|{\bf q''}|,\theta'',\theta')=\sum_{m=-1/2}^{1/2} \sum_{n=0}^{N} {\cal R}_{\lambda' \lambda'',n}^{\rho' \rho''}d_{\lambda' m}^{1/2}(\theta')d_{\lambda'' m}^{1/2}(\theta'')
 \int _{0}^{2 \pi }\frac{dx cos^{n} x}{d_{1}+d_{2}\cos
 x}e^{i(\lambda-m)x}.
\end{eqnarray}

\noindent After applying standard trigonometric manipulations:

\begin{widetext}
 \[ \cos^{2n-1} \theta = \frac{1}{2^{2n-2}}\left[ \cos(2n-1)\theta
 +\left( \begin{array}{c}
2n-1 \\
1 \end{array} \right) \cos(2n-3) \theta+
 ...+\left( \begin{array}{c}
2n-1 \\
n-1 \end{array} \right) \cos \theta \right], \] and
\[ \cos^{2n} \theta = \frac{1}{2^{2n}}\left( \begin{array}{c}
2n \\
n \end{array} \right)+\frac{1}{2^{2n-1}} \left[ \cos 2n \theta
 +\left( \begin{array}{c}
2n \\
1 \end{array} \right) \cos(2n-2) \theta+
 ...+\left( \begin{array}{c}
2n \\
n-1 \end{array} \right) \cos 2\theta \right]. \]

\end{widetext}
 the integral over the azimuthal angle of the
intermediate momentum in \Eqref{reducedv} can be reduced to
integrals of the following type:
\begin{eqnarray}
\eqlab{exactint} I_{m,n} = \int_0^{2\pi} \frac{d\phi \cos (m\phi)
e^{in\phi}}{1+ a \cos\phi} =
  \int_0^{2\pi} \frac{d\phi\cos (m\phi) \cos (n\phi)}{1+ a \cos
  \phi}.
\end{eqnarray}
\noindent For values $|a|<1$, this definite integral can be
evaluated analytically to obtain:
\begin{eqnarray}
\eqlab{intresult} I_{m,n}=\frac{\pi}{b}\left[
\left(\frac{b-1}{a}\right)^{m+n} +\left(\frac{b-1}{a}\right)^{|m-n|} \right]
\end{eqnarray}
where $b=\sqrt{1-a^2}$.

The results given above in  \Eqref{reducedv} work for all
standard particle exchanges in the s, t, or u channels.
Furthermore, it should be noted that additional azimuthal angle
dependences introduced by various form factors can easily be
handled by simple algebraic methods. The maximum power of $cos x$
needed for a particular diagram may increase (for example, N=2 for
u-channel $\Delta$ exchange).  In addition, $\bar{v}$ will, in
general, contain a sum of various terms corresponding to each
diagram included. However, all of these terms can be evaluated
using \Eqref{intresult}. In addition as noted earlier, this
procedure is not at all affected by the {\it equal time}
approximation and can be applied in the same manner to the full
4-D Bethe Salpeter equation.

\section*{Acknowledgments}
This work was performed in part under the auspices of the U. S.
Department of Energy, under the contract no. DE - FG02 -
93ER40756 with Ohio University and the National Science
Foundation under grant NSF - SGER - 0094668.


\end{document}